\documentclass[letterpaper, 10pt, conference]{ieeeconf}  % Comment this line out
\IEEEoverridecommandlockouts                              % This command is only
\usepackage[colorlinks=true,urlcolor=blue,linkcolor=blue]{hyperref} % navega por el doc
\usepackage{graphicx} % for pdf, bitmapped graphics files
\usepackage[ansinew]{inputenc} % Acepta caracteres en castellano
\usepackage{epstopdf}
\usepackage{tablefootnote}

\title{RACIMO@Bucaramanga: \\ 
A Citizen Science Project on Data Science and Climate Awareness}

\author{J. Pe\~na-Rodr\'iguez$^{1}$, P. A. Salgado-Meza$^{1}$, H. Asorey$^{2,3}$, L.A. N\'u\~nez$^{1,4}$, A. N\'u\~nez-Casti\~neyra$^{4}$,  \\ C. Sarmiento-Cano$^{1}$ and M. Su\'arez-Dur\'an$^{1}$ % <-this % stops a space
\thanks{$^{1}$Escuela de F\'isica, Universidad Industrial de Santander, Bucaramanga, Colombia.}
\thanks{$^{2}$Sede Andina, Universidad Nacional de R\'io Negro San Carlos de Bariloche, Argentina}
\thanks{$^{3}$ Laboratorio Detecci\'on de Part\'{\i}culas y 
Radiaci\'on, Instituto Balseiro y Centro At\'omico Bariloche, Comisi\'on Nacional de Energ\'{\i}a At\'omica, San Carlos de Bariloche, Argentina}
\thanks{$^{4}$Centro de F\'isica Fundamental, Universidad de Los Andes, 5101 M\'erida, Venezuela.}
}

\begin{document}
\maketitle
\thispagestyle{empty}
\pagestyle{empty}

%%%%%%%%%%%%%%%%%%%%%%%%%%%%%%%%%%%%%%%%%%%%%%%%%%%%%%%%%%%%%%%%%%%%%%%%%%%%%%%%
\begin{abstract}
This paper describes a collaborative experience to empower organized communities to produce, curate and disseminate environmental data. A particular emphasis is done on the description of open hardware \& software architecture and the processes of commissioning of the low cost Arduino-Raspberry-Pi weather station which measures: atmospheric pressure, temperature, humidity, precipitation, cloudiness, and illuminance/irradiance. The idea is to encourage more people to replicate this open-science initiative. 
We have started this experience training students \& teachers from seven mid secondary schools through a syllabus of 12 two-hours lectures with a web-based support which exposes them to basic concepts and practices of Citizen Science and Open Data Science. 
\end{abstract}
%%%%%%%%%%%%%%%%%%%%%%%%%%%%%%%%%%%%%%%%%%%%%%%%%%%%%%%%%%%%%%%%%%%%%%%%%%%%%%%%
\section{INTRODUCTION}
The impacts of climate change on urban living standards are currently one of the biggest concerns in our world\cite{IPCC2014}. Historically, monitoring climate variables relies mainly on official institutions which develop and support national and/or regional sensors networks. In Colombia, the Instituto de Hidrolog\'ia, Meteorolog\'ia y Estudios Ambientales (IDEAM\footnote{\url{http://www.ideam.gov.co/}}) has this national responsibility, and it is particularly noticeable its commitment to make all data publicly available\footnote{ \url{https://www.datos.gov.co}}.     

Despite these national efforts are trying to produce and disseminate useful and updated environmental data, increasing urban population, growing cities and cost-limitations, often means that such data-sets are not available at a range of spatio-temporal scales required to study extensive urban climate phenomena \cite{MullerEtal2015,KumarEtal2015,CostaResendeFernandes2015}. 

Nowadays a plethora of existing low cost Information and Communication Technology (ICT) devices, do-it-yourself (DIY) and do-it-with-others (DIWO) strategies, can increasingly augment the granularity and temporal resolution of the official monitoring networks.  More over, this confluence of low-cost-powerful devices and  people-centric strategies are transforming ordinary citizen into active producers, curators and data analyzers, understanding better their surrounding world, and taking part of the decision-making, to help improving their collective urban environment\cite{AggarwalAbdelzaher2013,SprakeRogers2014,NiforatosEtal2014}.

Emerging new paradigms of human-powered and people-centric sensing --Crowd Wisdom,  Collective Intelligence, Crowd-sourcing, Participatory Sensing, Crowd-sensing, among others-- are establishing and developing new approaches to produce, process, curate and disseminate a variety of urban environmental data.  (see \cite{CampbellEtal2008,DHondtStevensJacobs2013,GuoEtal2015,DelmastroArnaboldiConti2016} and references therein). 

To our knowledge, one of the most paradigmatic example of empowered communities with the confluence of DIY, DIWO, crowd-sourcing and participatory sensing was the Fukushima Daiichi Nuclear Power Plant grassroot radiation monitoring. There, the population of the Futaba District, supported from the Tokyo Hackerspace and by Reuseum\footnote{\url{http://reuseum.com}} --a company based on Boise, Idaho-USA closely related to US Hackerspace-- learned how to build Geiger counters and share data recorded from personal detectors. While official versions provide late information about  radiation levels from one of the most terrible nuclear incidents in the history of mankind, this empowered citizen network of sensors illustrates novel forms of decentralized disaster management and emergency responses \cite{KeraRodPeterova2013,Plantin2015}

Crowd-sourcing techniques are beginning to play crucial roles to record, preserve and disseminate urban environmental data, especially in densely populated and city governments are seriously considering these approaches as part of their participation policies and are encouraging  hackerspaces, FabLabs and university spin off to promote environmental citizen observatories (see \cite{MonimoPieraJurado2013,LanfranchiEtal2014,LiuEtal2014,WehnEvers2014,PalacinSilvaEtal2016}; and references therein). 

Several cities and regions around the world  --Amsterdam in the Netherlands\cite{JiangEtal2016}; Barcelona in Spain\cite{Diez2012}; Birminghan and Cambridge in United Kingdom\cite{ChapmanEtal2015,MeadEtal2013}; Ferrara region\cite{GravagnuoloLanzaraCovone2014} and Singapore\cite{KloecklEtal2011},  just to mention few of them-- are associated to environmental  quality initiatives through  citizen participatory sensing. 

This work describes RACIMO, (for its Spanish acronym of \textit{Red Ambiental Ciudadana de Monitoreo}, i.e. Environmental Citizen Monitoring Network) one of these experiences to empower organized communities to take decisions on environmental data produced and curated by themselves, which is taking place at Bucaramanga-Colombia. Particularly we will emphasize on the description of architecture (hardware and software) and the commissioning of the low cost arduino-Raspberry-Pi weather station which measures atmospheric pressure, temperature, humidity, precipitation, cloudiness, and illuminance/irradiance. The idea is to encourage more people to replicate this open-science initiative as much as possible.  

The present article is organized as follows: in the next section  we describe the project: its goals, its founding and its scopes. A particular attention is taken to describe the training effort to seven schools in Bucaramanga. Section \ref{UVAWeatherStation} presents the hardware and software architecture of the low cost weather station. In Section \ref{Commissioning} we give a detailed account of the calibration and commissioning process. This work ends with Section \ref{FutureDirections}, where we discuss the present status and future directions of the project.     

\section{RACIMO a citizen-science project}
\label{ProjectRacimo}
RACIMO is a collaborative experience to empower organized communities to produce, curate and disseminate environmental data. It was initially founded by the Regional Fund for Digital Innovation in Latin America and the Caribbean (FRIDA Program\footnote{\url{http://programafrida.net}} ) and the Vicerrectorado de Investigaci\'on y Extensi\'on de la Universidad Industrial de Santander Bucaramanga-Colombia.
 
\begin{figure}[!ht]
\begin{center}
\includegraphics[width=0.45\textwidth]{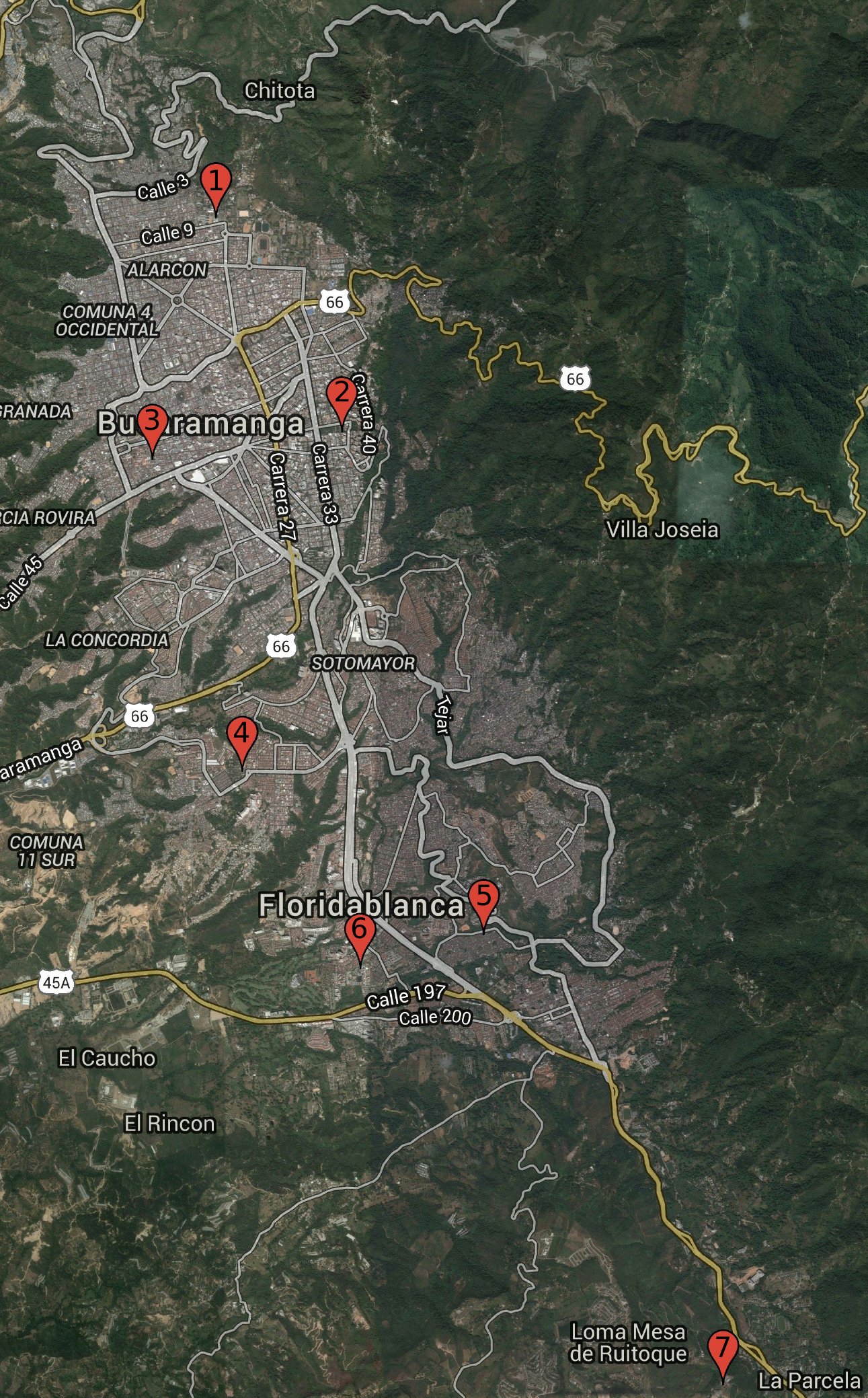}
\caption{RACIMO. UVAs location at high schools in the metropolitan zone of Bucaramanga. From the top to the bottom: 1) Universidad Industrial de Santander, 2) Pr\'incipe San Carlos, 3) Nuestra Se\~nora de F\'atima, 4) INEM, 5) Colegio Integrado Andes, CIANDES; 6) New Cambridge and 7) Saucar\'a. }
\label{Colegios}
\end{center}
\end{figure}

Because this project needs some basic expertise on data production and handling, we initiate it focusing on students of mid secondary school (grades 10 and 11) which usually are 15 or 16 years old\footnote{See a description of the education in Colombia \url{https://en.wikipedia.org/wiki/Education_in_Colombia}}. We started with 7 teams --one from each school-- made by a teachers and 4-5 students (see figure \ref{Colegios}) and we exposed them to basic concepts and practices of Citizen Science and Open Data Science. Through 12 sessions of 2 hours lectures we develop a syllabus covering: the emerging informational society and collective intelligence (4h); Basic statistics and data visualization (4h); Linux basics and Python programming  (4h); Environmental variables and sensors (4h); Basic electronics and Arduino configuration (4h); weather station configuration (4h). All these lectures have texts, short video-tutorials, assignments and related links, as it is displayed at the RACIMO website\footnote{\url{http://halley.uis.edu.co/tierra/}}.

\section{UVA: The low-cost weather station}
\label{UVAWeatherStation}
RACIMO is a network of automatic and autonomous weather stations of UVAs, (for its Spanish acronym of \textit{Unidades de Valoracion Ambiental} and both RACIMO and UVA are part of a worlplay, in Spanish UVAS are part of a RACIMO) with a variety of sensors which measures several environmental variables. Next sections will be devoted to describe the architecture (hardware \& software) of our low cost UVA.

\subsection{UVA building blocks}
\label{UVaBuidingBlock}
\begin{figure}[!ht]
\begin{center}
\includegraphics[width=0.3\textwidth]{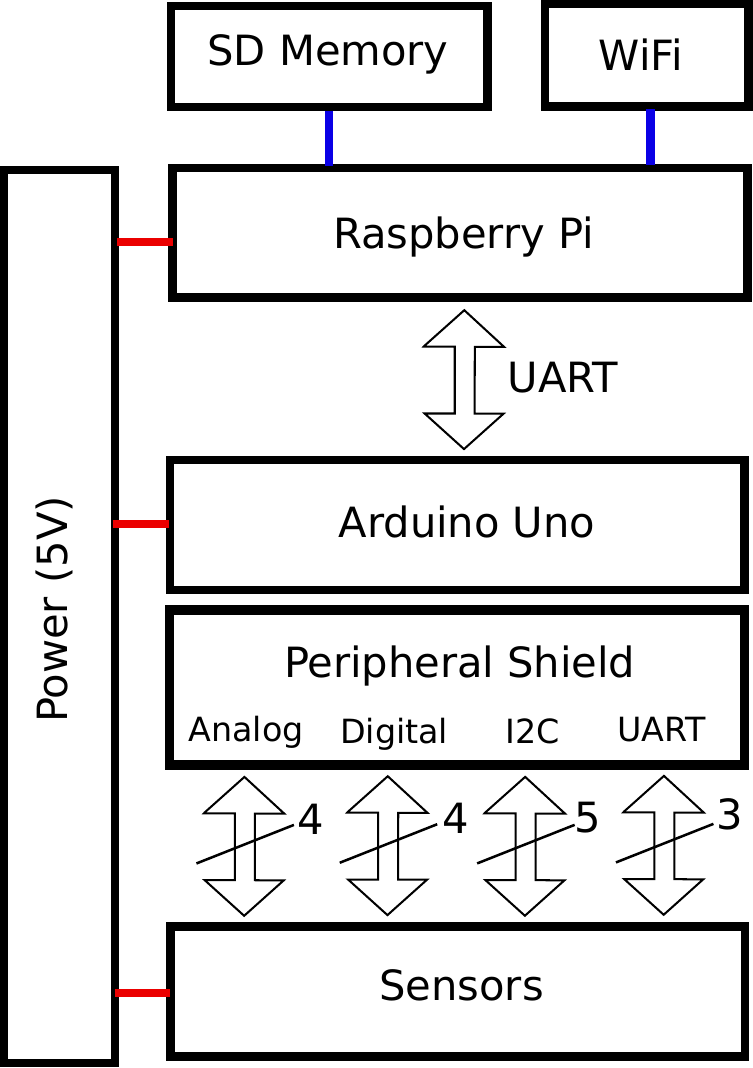}
\caption{UVA hardware schema. Sensors acquire and send information to the Arduino-Uno board through several protocols (I2C, UART or analog); then the Arduino-Uno creates a data packet which is transferred to the Raspberry-Pi by UART protocol. Finally, the Raspberry-Pi stores and sends the data to the repository. The entire UVA is powered by 5 volts to $\approx$ 600 mA. }
\label{UVA2}
\end{center}
\end{figure}

As it can be appreciated from fig \ref{UVA2}, UVA hardware architecture it is conceived having two main building blocks, embodied by an Arduino-Uno for sensing/pre-processing and, a Raspberry-Pi for data managing/storage.

The sensing block is represented by a peripheral shield where several basic sensors are connected to an Arduino-Uno board. It supports several protocols --such as: I2C, UART, analog and digital I/O-- and offers a cheap and flexible platform, to prepare data to be storage and/or sent to the next stage.  

The second UVA building block is an embedded linux single-board-computer (SBC) --a Raspberry-Pi 2 Model B-- which plays a central role managing, storing and sending the acquired data. It  was chosen due to its low cost, reduced power consumption, accessibility, great amount of information and resources for developers \cite{Moure2015}. In comparison with others SBCs, Raspberry-Pi has a better performance/price relation \footnote{\url{https://www.cooking-hacks.com/blog/new-linux-embedded-devices-comparison-arduino-beagleboard-rascal-raspberry-pi-cubieboard-and-pcduino/}}. The main features of our SBC are: a RAM of 1 GB, a 900MHz quad-core ARM Cortex-A7 CPU, 10/100 Mbps Ethernet, and supports a SD memory until 128 GB\footnote{\url{http://elinux.org/RPi_SD_cards}}, which could store data of an UVA during 4 years taking in account that data are recorded with a frequency of sampling of 1 Hz.

\subsection{UVA sensors}
\label{UVASensors}
In our weather station we can distinguish two type of sensors: \textit{regular} and \textit{software-enhanced}. In this section we shall describe our regular or conventional sensors and, later we will discuss what we have called software-enhanced ones. 

Among the regular ones can differentiate commercial and in-house-devised sensors. These later devices are implemented based on particular gadgets calibrated to gauge indirectly the specific variables we are interested to measured. Software-enhanced sensors are those which needs significant amount of computation to infer a particular variable to be considered.

\begin{table}[ht]
\caption{UVA sensors. Regular commercial: MPL115A2, HIH-4030, and RG-11. In-house-devised: illuminance, and Irradiance.; Software-enhanced: Cloudiness}
\label{Table1}
\begin{center}
    \begin{tabular}{ | l | l | l | l |}
    \hline
    \textbf{Sensor} & \textbf{Variable} & \textbf{Accuracy} & \textbf{Units} \\ \hline
    MPL115A2& Pressure and T$^{\circ}$ & $\pm$1kPa, $\pm$0.18$^{\circ}$C  & hPa and $^{\circ}$C \\ \hline
    HIH4030 & Humidity & $\pm$3.5 $\%$RH & $\%$RH \\ \hline
	RG11 & Rain  & 0.069 mm & mm \\ \hline    
    LDR & Illuminance & 127 Lux  & Lux \\ \hline
		Solar cell & Irradiance & 1.2 W/$m^2$ & W$/m^2$ \\ \hline
		Raspicam & Cloudiness & NA & $ \% $ \\
		\hline
    \end{tabular}    
\end{center}
\end{table}

\subsubsection{UVA regular sensors}
\label{UVADirectSensors}
Currently, our UVA has 6 direct sensors which are described in Table \ref{Table1} and some of these basic sensors are standard low cost commercial ``plug \& play'' devices: 
\begin{itemize}
\item \textbf{MPL115A2} is an absolute pressure sensor with a digital I2C output for low cost applications;
\item \textbf{HIH-4030} measures relative humidity (\%RH) and delivers it as an analog output voltage;
\item \textbf{RG-11} is a robust and sensitive optical rain gauge which senses water hitting on its outside surface by beams of infrared light.
\end{itemize}
Here ``plug \& play'' means that these sensors have a (almost) direct calibration protocol.   

Our in-house-devised sensors are:
\begin{itemize}
\item \textbf{Illuminance} sensor is based on a Light Dependent Resistor (LDR), which decreases its resistance with increasing incident light intensity;
\item \textbf{Irradiance} detector uses a cheap small (15 x 10 cm) solar cell which converts solar energy directly into electrical current by the photo-voltaic effect.
% \item \textbf{Environmental noise} sensor was designed with a standard inexpensive  microphone  which converts sound in a electrical signal, then this signal is digitalized and preconditioned for calculate the sound power level.
\end{itemize}
These commercial and in-house-devised sensors were calibrated in our laboratory and details are included below, in section \ref{Commissioning}.

\subsubsection{UVA software-enhanced sensors}
\label{UVAEnhanceSensors}
\begin{figure}[!ht]
\begin{center}
\includegraphics[width=0.25\textwidth]{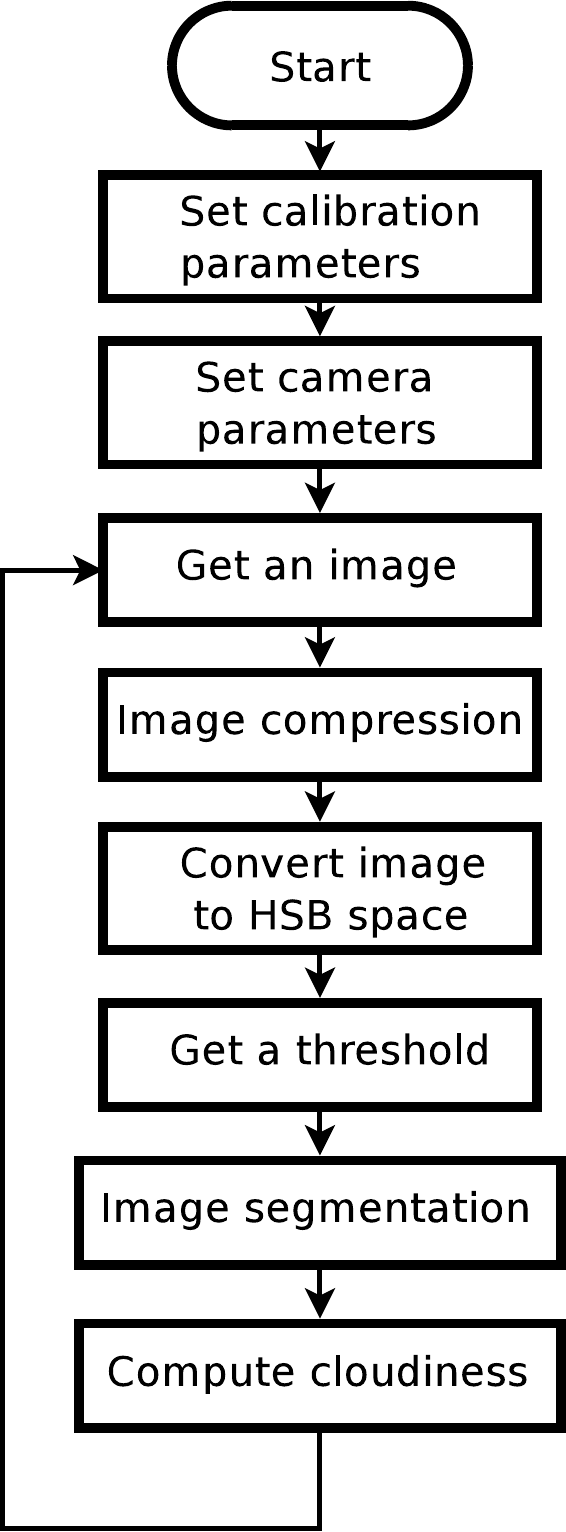}
\caption{Cloudiness algorithm. The algorithm to calculate cloudiness is based on an image segmentation task. That is, a threshold is applied on the image depending of its brightness.}
\label{Cloudiness}
\end{center}
\end{figure}

Our cloudiness sensor is also in-house-devised based on a \textbf{Raspicam}  -- a small size and low consumption camera,  capable to take 2592 x 1944 pixel static images--  attached to the Raspberry-Pi. Images are processed with an implementation of a segmentation-based algorithm, using Adaptative Average Brightness Thresholding (AABT). A simple algorithm, with good performance and low computational cost\cite{Leung1995}. Because it spends around 20 seconds to estimate the cloudiness, data are temporary stored emulating a sample and hold system. A flowchart of this implementation is shown in figure \ref{Cloudiness}.

\subsection{UVA data flow}
\label{UVADataFlow}
UVA data flow, is fully automatic and starts with the Arduino data capture with regular sensors, with this information the raw data packet is built and sent to the the Rapsberry-Pi where the cloudiness data is added. Next, the corresponding metadata is attached to the data file (see figure \ref{DataFlow} for an illustration). The data is recorded with a sampling frequency of 1 Hz, the time of sampling is self-adapted depending on the number of sensors connected.  

The structure of this initial data-packet is: UTC time of position (hhmmss.sss), atmospheric pressure, temperature, humidity, irradiance, illuminance, pluviosity, latitude and longitude.

Then, the Raspberry attaches two other values (UTC time, in Unix format and cloudiness) and finally the data-record is built. Each item starts with an alphanumerical sign (\#) to distinguish it from the metadata information.  

The data-file is assemble with a sequence of data-records and the appropriate metadata. The metadata information provides information about the status of each UVA, i.e. : UVA name, firmware version, location, sensors connected, power supply  (grid or photo-voltaic), GPS, time zone and data structure. When a sensor is not connected, a value of -666 is placed to keep the structure of the data file. The syntax to name of data-files is: Data\_year\_month\_day\_hour.txt. 

\begin{figure}[!ht]
\begin{center}
\includegraphics[width=0.4\textwidth]{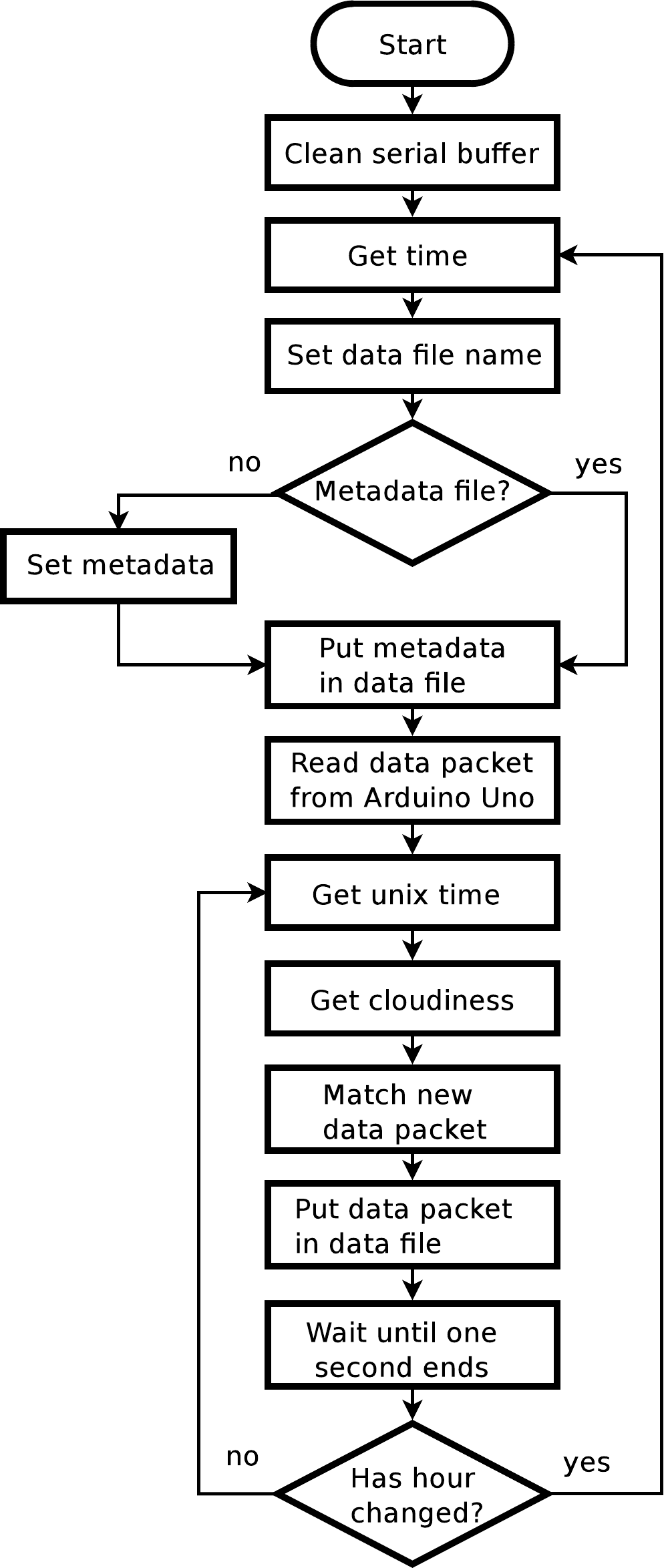}
\caption{Data recording algorithm. The algorithm creates data files composed by meta-data and data provided for sensors. The name of each data file changes hourly. The dashed line indicates tasks done outer the Raspberry-Pi}
\label{DataRecording}
\end{center}
\end{figure}

\begin{figure}[!ht]
\begin{center}
\includegraphics[width=0.5\textwidth]{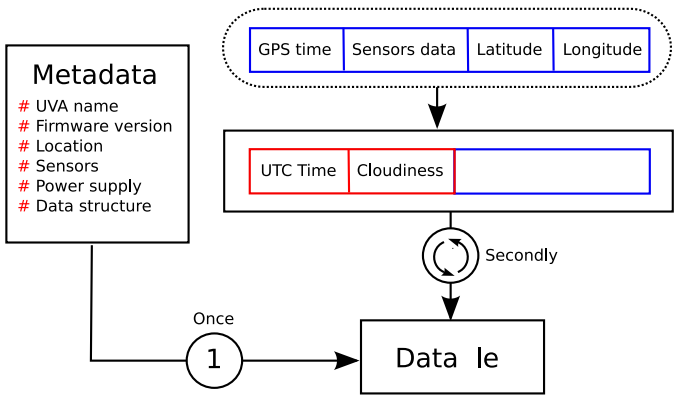}
\caption{Data flow schema. Dashed line enclose the data recording process made by the Arduino-Uno board. Generation of the datafile is made in the Raspberry-Pi, firstly, the meta-data is set in the file, then a data packet, which contain sensor information, is write secondly.}
\label{DataFlow}
\end{center}
\end{figure}

We have implemented a \textit{watchdog} system on a \textit{cron (Unix)} to manage tasks for testing the recording process at different stages. This implementation is part of a set of actions to secure the autonomy and confidence of the UVA operation. These actions are checked with various frequencies, i.e., the recording process is verified each minute, the WiFi connection each 5 minutes and data is uploaded each 24 hours. Besides, if the power supply of the UVA turns off, we have developed a checkpoint process to restart all tasks automatically without any external intervention.

\section{Commissioning and Calibration}
\label{Commissioning}
In this section we shall describe the calibration processes for all sensors described above in table \ref{Table1}.

\begin{figure}[!ht]
\begin{center}
\includegraphics[width=0.45\textwidth]{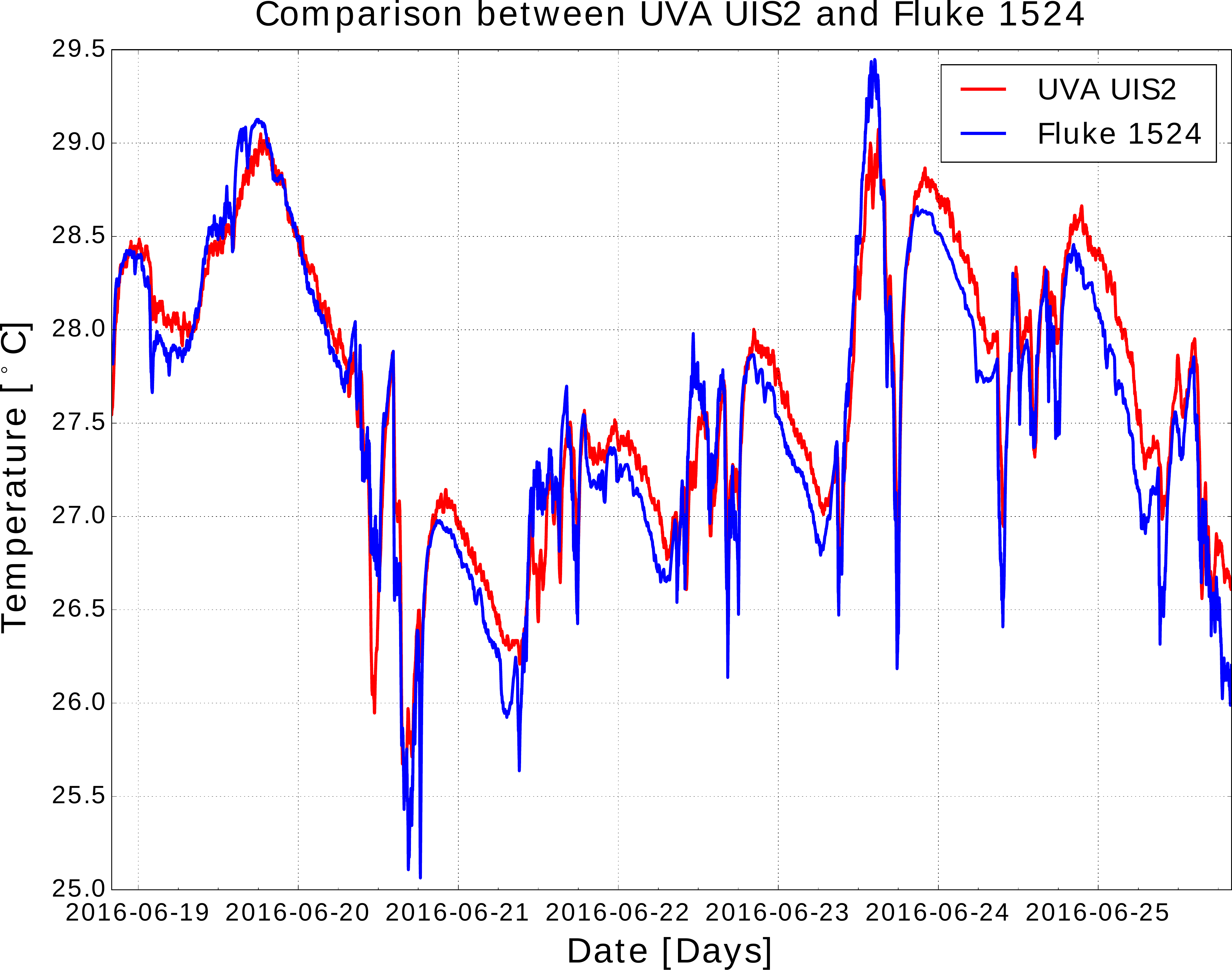}
\caption{Test of the temperature sensor of UVA UIS2. Data were recorded, at room temperature during a week --June 19-25, 2016--  UVA UIS2 (red line) and the Fluke 1524 Temperature Meter (blue line). The UVA UIS2 data was processed for a mean filter with a time window of 51 seconds. The difference between data from UVA UIS2 and the Fluke 1524 Meter is overall under 1 $^{\circ}$C.}
\label{Fluke1}
\end{center}
\end{figure}

The \textbf{MPL115A2} senses temperature and pressure and comes pre-calibrated\footnote{\url{http://www.nxp.com/assets/documents/data/en/data-sheets/MPL115A2.pdf}}. To test the quality of temperature pre-calibration we use Fluke 1524 Temperature Meter, which has an accuracy of $\pm$ 0.24 $^{\circ}$C. In Fig. \ref{Fluke1} it is shown the comparison both sensors, under the same environmental conditions, during a week of data recording. It is clear that  
difference between data from UVA UIS2 and the Fluke 1524 Meter is overall under 1 $^{\circ}$C. %The histogram of the temperature difference, with 604800 samples, is shown in Fig. \ref{histoTemp}.

% \begin{figure}[!ht]
% \begin{center}
% \includegraphics[width=0.45\textwidth]{Figures/hist2}
% \caption{Histogram of the difference between the temperature sensor of UVA UIS2 and the Fluke 1524 Temperature Meter. Near of the 50\% of the values are under 0.5$^{\circ}$C therefore the bias is not significant.}
% \label{histoTemp}
% \end{center}
% \end{figure}

Humidity sensor \textbf{HIH4030} calibration was made using the information of the ratio between output voltage and relative humidity presented in its datasheet\footnote{\url{https://www.sparkfun.com/datasheets/Sensors/Weather/SEN-09569-HIH-4030-datasheet.pdf}}. 

\begin{figure}[!ht]
\begin{center}
\includegraphics[width=0.45\textwidth]{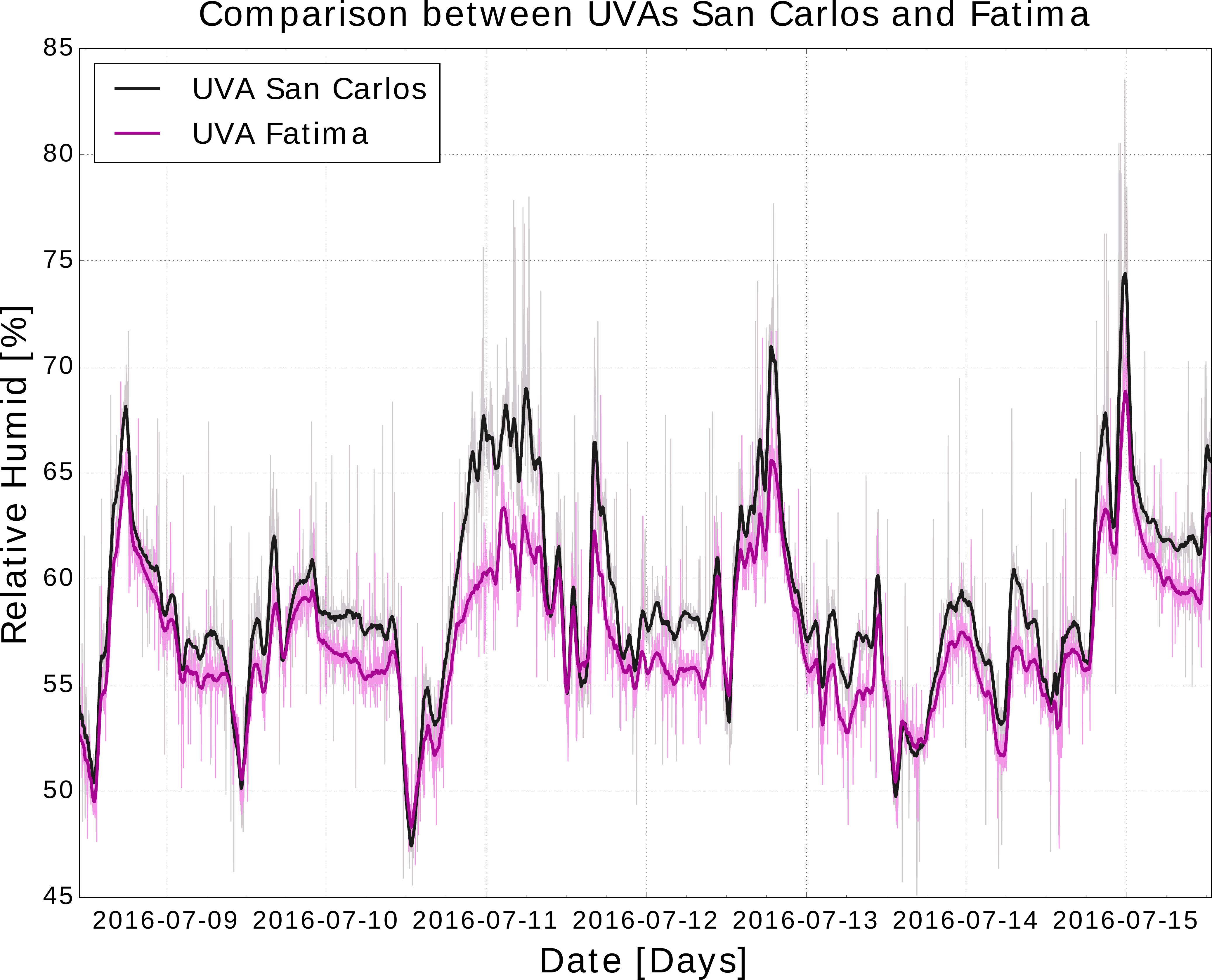}
\caption{Humidity comparison between UVAs San Carlos (black line) and Fatima(magenta line). Data were recorded during a week, July 9-15, 2016. UVAs were placed in the same room. Data of both UVAs were processed for a mean filter with a time window of 51 seconds. Blurred lines behind signals are raw data.}
\label{Humidity}
\end{center}
\end{figure}

A comparison between two different UVAs --Fatima and San Carlos-- after calibration stage is shown in Fig. \ref{Humidity}. The maximum variability between both UVAs was 5\% with data recorded during a week (July 9-15) and both UVAs were placed in the same room to guarantee equivalent environmental conditions. 

\begin{figure}[!ht]
\begin{center}
\includegraphics[width=0.5\textwidth]{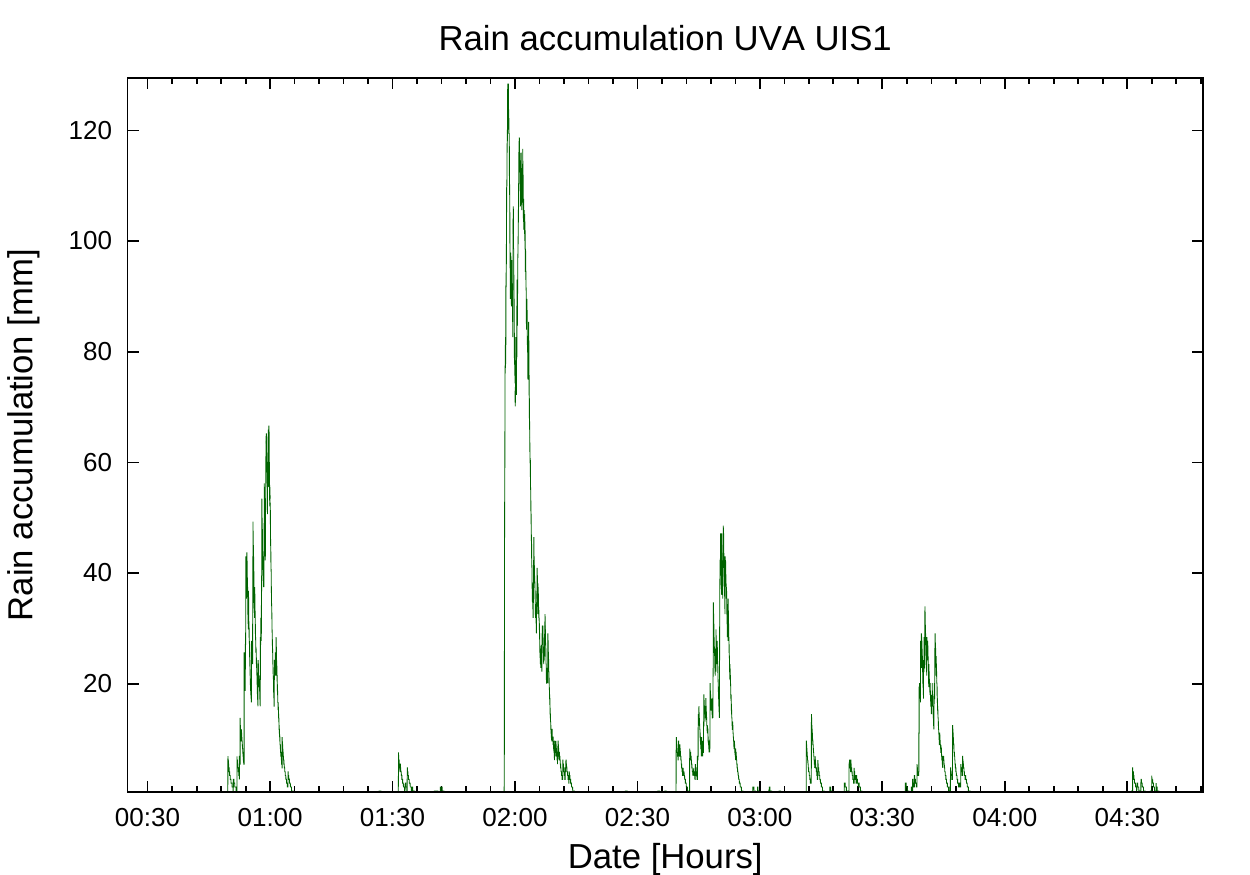}
\caption{Rain accumulation recorded for UVA UIS1 on April 12. Rain transient events produced during a storm are clearly shown one of them having a maximum at 2 a.m. whose accumulation reached over 120 m.m.}
\label{Rain}
\end{center}
\end{figure}

The \textbf{RG11} has an accumulation rain register of 8 bits, having a maximum value (255) which can be associated, from the manufacturing calibration curve, to 0.7 inches or 17,78 mm of rain\footnote{\url{http://rainsensors.com/rain-gauge-modes/}}. This sensor is capable to detect transient event in short times, as it is illustrated in figure \ref{Rain}, for data recorded on April 12.

A major advantage of UVAs is that they do not have mechanical parts in all its sensors, making it more robust and minimizing mechanical wear incident. This is particularly important in the rain gauge sensor, RG11, which differs of typical pluviometers that use a small bucket which is emptying continuously, in this case, it measures the drops accumulation during a storm by means of an optical system. 

%%%
Calibration of the \textbf{LDR} was made using the resistance-illumination curve, which can be expressed as,
\begin{equation}
E=AR^{-\frac{1}{B}}
\end{equation}
where, $E$ is the illuminance, $R$ the resistance and $A,B$ are constants resulting of the curve parameterization.

The \textbf{irradiance}, $G$ can be obtained by modeling the solar cell behaviour as follows,
\begin{equation}
G = \frac{IV}{nS}
\end{equation}
where $I$ is the output current, $V$ the output voltage, $n$ the efficiency, and $S$ the sensitive area of the solar cell. 

\begin{figure}[!ht]
\begin{center}
\includegraphics[width=0.5\textwidth]{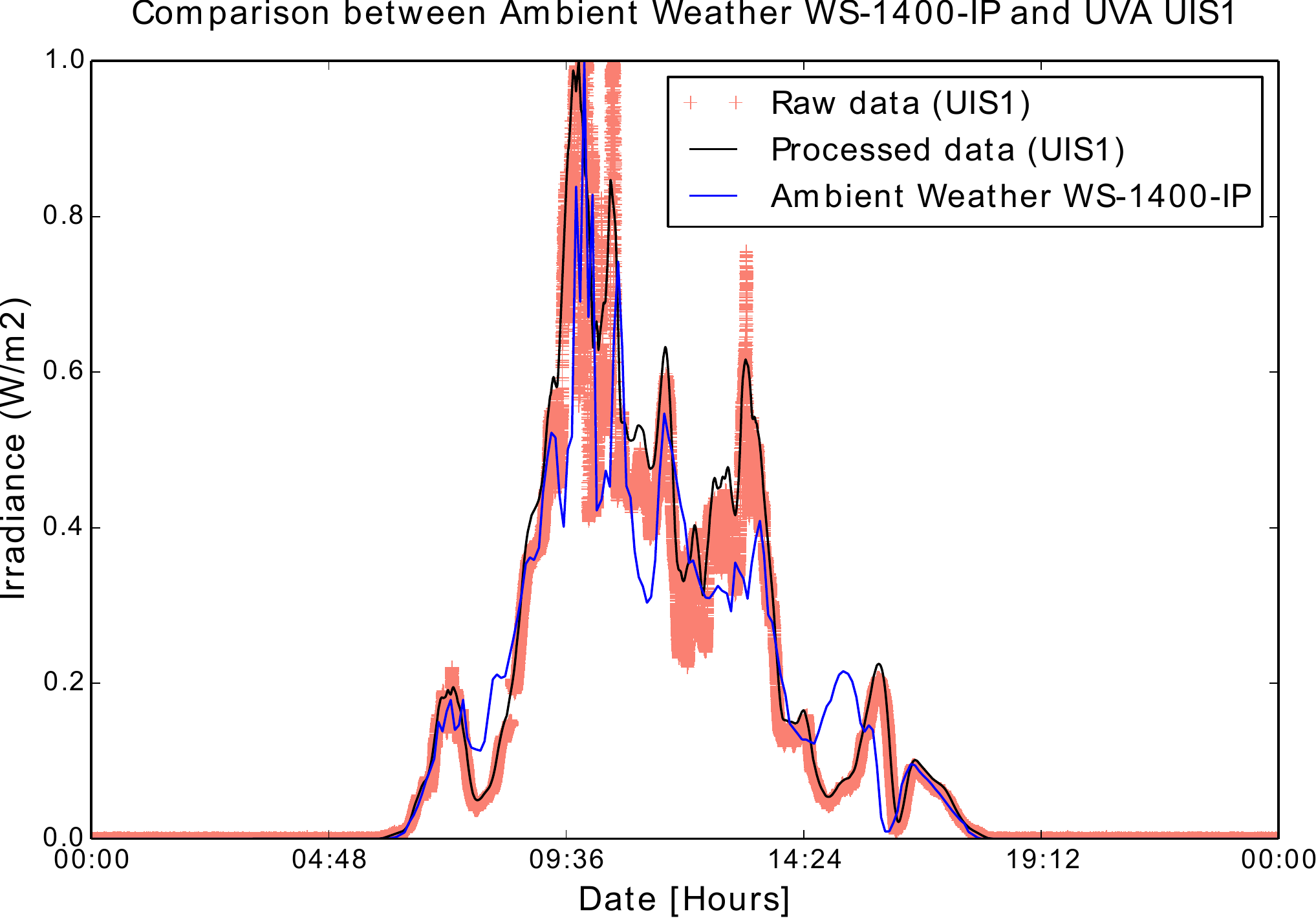}
\caption{Irradiance measurement comparison between UVA and an Ambient Weather WS-1400-IP (blue line) on July 16, 2016. Raw data (dashed line) and the processed data (black line) of UVA are shown. Data was processed through a mean filter with a time window of 1001 seconds.}
\label{irradiance}
\end{center}
\end{figure}

The calibration was carried out by comparing the UVA irradiance measurements with  data provided for an Ambient Weather WS-1400-IP (pattern station\footnote{\url{http://www.ambientweather.com/amws1400ip.html}}). The process of calibration consisted of correcting the bias between the UVA and the pattern data. Fig. \ref{irradiance} shows the measured irradiance comparison between UVA UIS1 and the pattern station.

% * <lnunez@uis.edu.co> 2017-04-27T22:08:36.657Z:
% 
% Primero, no se entiende el cuento del bias que fue corregido. Cómo fue el proceso para corregirlo y crear (creo) los datos procesados. Ese filtro de que hablas en la figura como fue el cuento
% Segundo hace falta algún criterio estadístico que nos refuerce los dibujitos Esto es que muestre que los datos corregidos
% para la UAV se parecen mucho a los de la estación de calibración. 
% La función es la misma pero hay un bias que se añade
% ^.

% \textbf{Noise} level, $L$, can be measured from variations of a ``reference pressure'' as 
% \begin{equation}
% L=20 \log_{10}\frac{P_{RMS}}{P_{ref}}
% \end{equation}where, $P_{ref}$ is the reference pressure ($20$x$10^{-6} Pa$) and $P_{RMS}$ is the \textit{Root Mean Square} pressure measured through the microphone depending of a relation between its output voltage and efficiency (calibration curve).

\begin{figure}[!ht]
\begin{center}
\includegraphics[width=0.5\textwidth]{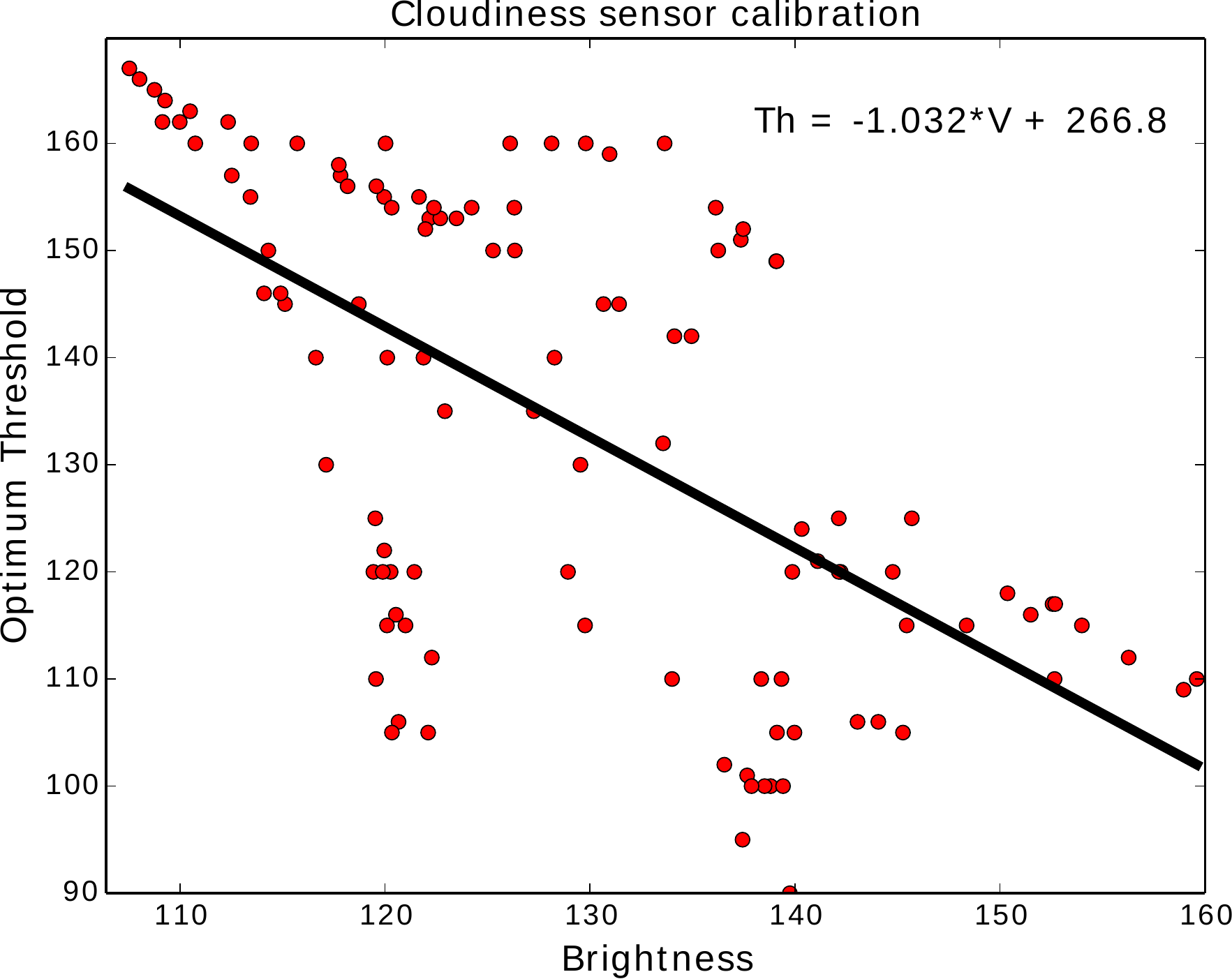}
\caption{Cloudiness calibration. The model was found by means a linear regression method applied on 108 samples, each sample is the brightness (V) and the optimum segmentation threshold (Th) of a random image took by a UVA. The linear function of the model is $Th = -1.032*V + 266.8$.}
\label{Figure6}
\end{center}
\end{figure}

\begin{figure}[!ht]
\begin{center}
\includegraphics[width=0.4\textwidth]{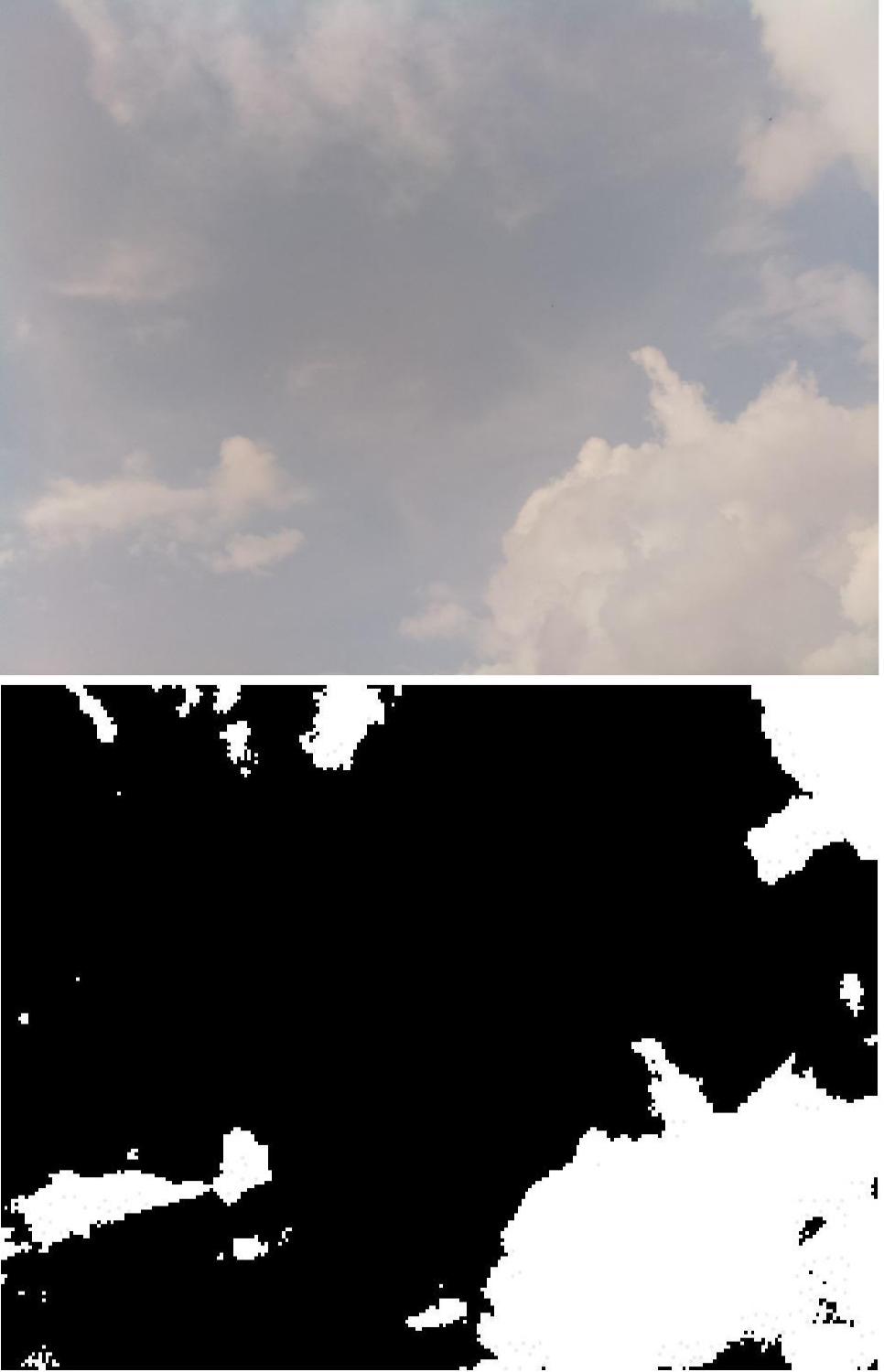}
\caption{Clouds segmentation example. At the top, the raw image. At the bottom, the image segmented by means a threshold depending of brightness.}
\label{Figure7}
\end{center}
\end{figure}

Calibration of \textbf{cloudiness} sensor was made by using 108 images of a partially cloudy sky. For each image, we set the optimum threshold at same time the image brightness is calculated. Then, we fit a model that predicts the segmentation threshold depending of brightness. In the Fig. \ref{Figure6} shows the results of the calibration process and the model fitting. A cloudiness segmentation example resulting after the calibration process is shown in Fig. \ref{Figure7}.
% * <lnunez@uis.edu.co> 2017-04-27T23:23:41.984Z:
% 
% No se entiende nada del modo de la calibración de la nubosidad. Pero nada es NADA.  Por favor hazla completa otra vez.
% 
% ^.

\begin{figure}[!ht]
\begin{center}
\includegraphics[width=0.5\textwidth]{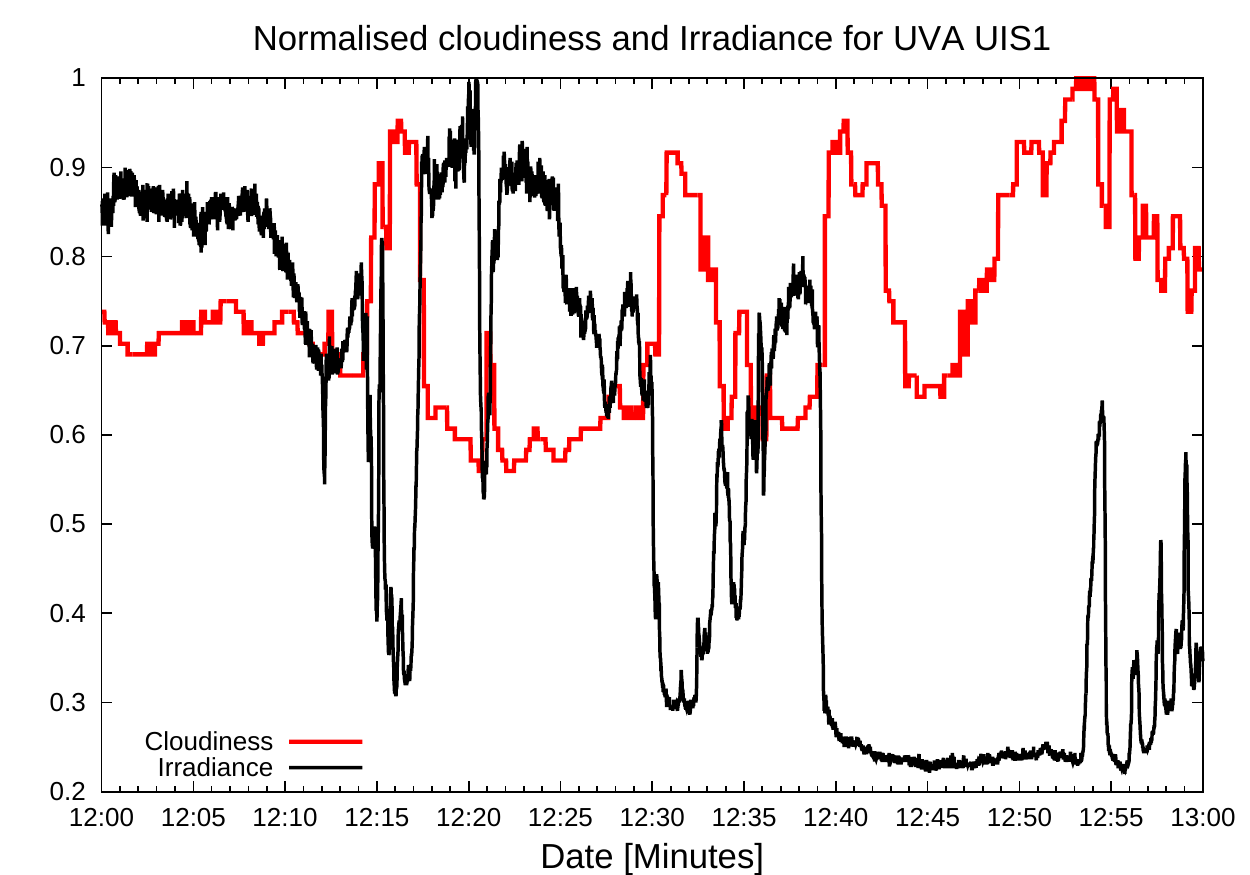}
\caption{Comparison between irradiance and cloudiness measured by UVA UIS1 at midday on August 1, 2016. The signals present an inversely proportional relation which can be quantify through Pearson correlation coefficient, in this case is -0.67.}
\label{CloudinessIrradiance}
\end{center}
\end{figure}

Next, a anti correlation test for cloudiness and irradiance sensors was carried out through a comparison between both dataset recorded during one hour. 
The anti-correlated relation between irradiance and cloudiness can also be used  for developing a model to predict cloud covering\cite{Luo2010}.
Fig. \ref{CloudinessIrradiance} presents a normalized comparison between cloudiness and radiation measured by the UVA UIS1. Cloudiness behaves inversely to irradiance, thus when cloudiness increases the solar radiation decreases.

%  Nowadays, air environmental chemical quality has become a big concern in cities around the world. RACIMO has the capability to measure some of the most important air chemical pollutant. Currently, each UVA has a CO2 and a NO2 sensor, they can measure the behavior of these variables along the time. In the Fig. \ref{Figure14} CO2 concentration measures during four days, July 9-12, are shown. Data have a day-night modulation which is been studied to know its source. 
% * <lnunez@uis.edu.co> 2017-04-27T23:41:54.742Z:
% 
% Como se realizó la calibración de estos sensores de gas? No se dice en ningún lado.
% no se han calibrado, quitar
% ^.

% \begin{figure}[!ht]
% \begin{center}
% \includegraphics[width=0.5\textwidth]{Figures/CO2_week}
% \caption{Measurement of CO2 concentration in the UVA UIS1 during four days, July 9-12, 2016. Data show a day-night modulated behavior which has the maximum at midnight and the minimum at midday. The difference between them is around 1 ppm. }
% \label{Figure14}
% \end{center}
% \end{figure}

The energy consumption of a UVAs was measured in several times during a day, 100 samples of current and voltage values were recorded. The current average was 623\,mA and the voltage was 5.23\,V, the mean of power consumption was 3.26\,W. 

% * <lnunez@uis.edu.co> 2017-04-27T23:46:54.740Z:
% 
% Aquí sería bueno especificar la superficie de los paneles que se requiere.
% 
% ^.se puede alimentar con un panel solar de 20 W de 52 cm x 36 cm

\section{Status and future directions}
\label{FutureDirections}
We have presented whole RACIMO project, i.e. its conceptual foundations; its relation to other citizen science projects; available online training resources; the weather station hardware/software architecture and its calibration/commissioning . 

We have described our totally automatic and low consumption weather station built embracing the open hardware/software paradigms which allow users to easily extend its the capabilities, attaching new sensors and algorithms. Our first model has showed high performance and robustness when recording processes during several weather conditions.

The performance of the AABT algorithm was evaluated through a Chi-square test ($\chi^2$) and a Mean Square Error (MSE) value. The model was fitted using 110 samples and the validation results show a $\chi^2=241$ and MSE$=293$, which indicates although the algorithm has an acceptable performance, it should be improved. As a future work, a model fitting increasing the samples population has been proposed taking in account the clear dependency between the segmentation threshold and the brightness.

\addtolength{\textheight}{-12cm}   % This command serves to balance the column lengths

%%%%%%%%%%%%%%%%%%%%%%%%%%%%%%%%%%%%%%%%%%%%%%%%%%%%%%%%%%%%%%%%%%%%%%%%%%%%%%%%

%%%%%%%%%%%%%%%%%%%%%%%%%%%%%%%%%%%%%%%%%%%%%%%%%%%%%%%%%%%%%%%%%%%%%%%%%%%%%%%%



\section*{ACKNOWLEDGMENT}
We gratefully acknowledge  the financial support from the Regional Fund for Digital Innovation in Latin America and the Caribbean (FRIDA) under project 347 and  Vicerrectorado de Investigaci\'on y Extensi\'on de la Universidad de Santander Bucaramanga-Colombia.  We also thanks the enthusiasm and dedication of the teachers and students of  six high schools in Bucaramanga: INEM, Nuestra Se\~nora de F\'atima, Pr\'incipe San Carlos, New Cambridge, CIANDES and Saucar\'a.

\bibliographystyle{unsrt}
\bibliography{BibTex/BiblioRACIMO.bib}

\end{document}